\documentclass[aps,prx,reprint,superscriptaddress]{revtex4-1}

\usepackage[version=3]{mhchem} 
\usepackage{graphicx} 
\newcommand{\angstrom}{\mbox{\normalfont\AA}}
\usepackage{braket}
\usepackage{bm} 

\begin{document}


\title{Quantum plasmonic nanoantennas}


\author{Jamie M. Fitzgerald}
\email{jf1914@ic.ac.uk}
\affiliation{Department of Physics, Condensed Matter Theory, Imperial College London, London SW7 2AZ, United Kingdom}
\author{Sam Azadi}
\affiliation{Department of Materials Science, Royal School of Mines, Thomas Young Center, London Centre
for Nanotechnology, Imperial College London, London SW7 2AZ, United Kingdom}
\author{Vincenzo Giannini}
\homepage{www.gianninilab.com}
\affiliation{Department of Physics, Condensed Matter Theory, Imperial College London, London SW7 2AZ, United Kingdom}


\date{\today}

\begin{abstract}
We study plasmonic excitations in the limit of few electrons, in one-atom thick sodium chains, and characterize them based on collectivity. We also compare the excitations to classical localised plasmon modes and find for the longitudinal mode a quantum-classical transition around 10 atoms. The transverse mode appears at much higher energies than predicted classically for all chain lengths. The electric field enhancement is also considered which is made possible by considering the effects of electron-phonon coupling on the broadening of the electronic spectra. Large field enhancements are possible on the molecular level allowing us to consider the validity of using molecules as the ultimate small size limit of plasmonic antennas. Additionally, we consider the case of a dimer system of two sodium chains, where the gap can be considered as a picocavity, and we analyse the charge-transfer states and their dependence on the gap size as well as chain size. Our results and methods are useful for understanding and developing ultra-small, tunable and novel plasmonic devices that utilise quantum effects that could have applications in quantum optics, quantum metamaterials, cavity-quantum electrodynamics and controlling chemical reactions, as well as deepening our understanding of localised plasmons in low dimensional molecular systems.
\end{abstract}


\maketitle

\section{Introduction}
Localised surface plasmons (LSPs) are collective oscillations in finite free-carrier systems driven by electromagnetic fields which dominate the optical response of metals near the plasmon frequency \cite{Maier2007}. Classically, they can be understood as a resonance arising from the restoring force of induced surface charges and, as such, are highly sensitive to the geometry and surrounding environment and lead to strong electric field confinement and enhancement which results in a wealth of potential applications such as sensing \cite{Homola1999}, metamaterials/metasurfaces \cite{Boltasseva2011}, colour generation \cite{Kristensen2016}, controlling chemical reactions \cite{Yan2016} and energy conversion \cite{Atwater2010}.\\
\indent In quantum plasmonics \cite{Fitzgerald2016}, one aims to describe how the constituent electrons behave under a perturbation to form a collective mode. This is an important task in modern plasmonics as recent state-of-the-art experiments are exploring size regimes, such as extremely small metallic nanoparticles (MNPs) \cite{Scholl2012} and sub-nanometer gaps, \cite{Ciraci2012,Savage2012,Scholl2013,Scholl2015} where classical laws, which have been used successfully in the vast majority of plasmonic research done to date, will fail. In fact, even quasi-quantum mechanical theories, like the hydrodynamic model \cite{Raza2011,Mortensen2014}, may not be adequate for quantitative predictions \cite{Stella2013}. Instead the electrons must be treated using \emph{ab initio} methods such as time dependent density functional theory (TDDFT). This will automatically take into account modification of the optical response from electron spill out, energy level discreteness and non-locality: which arises as a consequence of the finite extent of the electronic wavefunction. It can also be important to explicitly include the underlying atomic structure rather than relying on simplifications such as the jellium model \cite{Zhang2014}. In this picture, MNPs behave as large molecules and the plasmon, which captures much of the spectral weight in the extinction cross-section, is built up out of transitions between discrete energy levels mediated by the Coulomb interaction.\\
\indent In this context it is important to have clear and robust methods to analyse the results from calculations which may return a large amount of data and many excitations over the frequency range of interest. The surface of the MNP breaks translational invariance and allows mixing of plasmons and single particle excitations (Landau damping) meaning identification of peaks can be difficult. Often any peaks with a large dipole strength are claimed to be plasmonic \cite{Gao2005,Yan2007,Yan2008} using the argument that this indicates collectivity. This can be erroneous as the dipole strength function, as the name suggests, simply gives a measure of the dipole moment and hence how strongly the excitation interacts with light, a single-electron hole pair separated over a large distance can give a large dipole moment but it does not seem sensible to describe it as plasmonic! Thus to identify plasmons in small clusters more sophisticated methods of identification are needed. Examples in the literature include analysis of the electron population dynamics \cite{Townsend2014,Ma2015}, the Coulomb scaling method \cite{Bernadotte2013} and the plasmonicity index \cite{Bursi2016}. Collectivity is one of the most intuitive characteristics of a plasmon, in this work we focus on measuring collectivity to identify plasmonic excitations. Using excited state structural analysis to quantify the collectivity of an excitation, which can be obtained from the standard output of any electronic structure code (as long as the transition density matrix can be calculated), we define a new index which is the product of the collectivity index \cite{Martin2003,Plasser2014} and the dipole strength. This provides a simple and convenient method to quantify how plasmon-like a molecular excitation is.\\
\begin{figure*}
\includegraphics[width=17.2 cm]{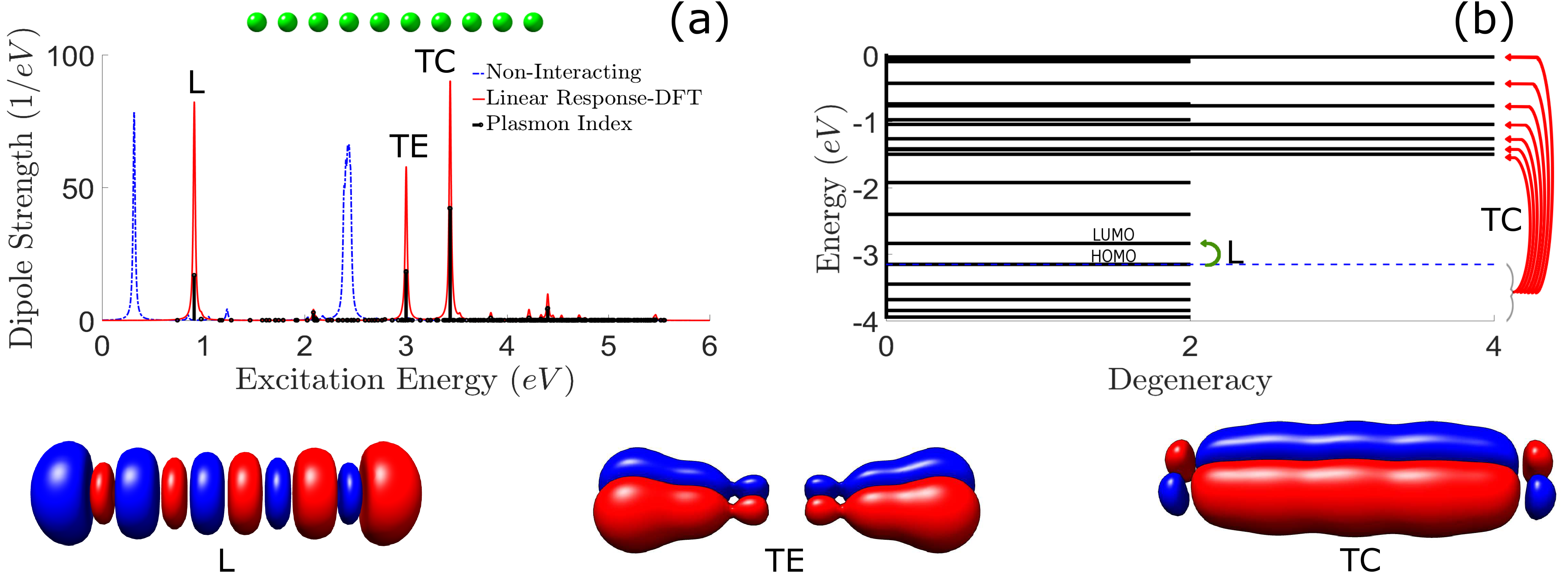}
\caption{(a) Dipole strength of a \ce{Na10} chain with the transition densities for the longitudinal (L) and transverse end (TE) and central (TC) modes shown. The vertical black bars indicate the collectivity index multiplied by the oscillator strength for every excitation and reveal the most plasmonic peaks. The 3D transition density plots are created with UCSF Chimera \cite{Pettersen2004}. (b) Visualisation of the transitions that make up the L and TC excitation.} \label{fig:figure_1}
\end{figure*}
\indent One of the most widely studied plasmonics systems are metallic nanoantennas \cite{Muskens2007,Biagioni2012,Giannini2011} where oscillation of free carriers in a confined space allows the reception or emission of electromagnetic radiation. To achieve a large optical response in the visible one needs to tune the resonant frequency by shrinking the antenna down to smaller sizes. Nanorods, in particular, have been studied in depth and, within the quasi-static limit ($\lambda \gg L,R$, where $L$ is the length of the nanorod and $R$ is the radius), the plasmon resonance depends only on the aspect ratio $L/R$ \cite{Bryant2008}. They are also one of the simplest anisotropic structures which lead to strong and tunable resonances. Single atomic thick atomic chains, which have become a standard `workhorse' of quantum plasmonics \cite{Gao2005,Yan2007,Yan2008,Yasuike2011,Bernadotte2013,Piccini2013}, represent the ultimate aspect ratio limit of nanorods. Their small size could lead to added tunability and novel quantum plasmonic behaviour as well as aid in our understanding of how plasmons arise in larger systems within a quantum picture. Nanorods have also been explored within the jellium approximation using TDDFT \cite{Zuloaga2010}. Using linear response TDDFT (a.k.a.\ the Casida method) on the open source code OCTOPUS \cite{Andrade2015} we find that while the electronic excitations of such chains may share some characteristics with classical plasmons, \emph{the longitudinal mode in particular can be well described by a classical ellipsoid model above a threshold of about 10 atoms. In other respects, they differ considerably with the transverse mode appearing at much higher energies.} This reveals that the atomic chain is a very quantum system no matter how long the chain is made.\\
\indent One of the most crucial physical quantities to consider in plasmonics is the field enhancement. Unfortunately this is difficult to calculate in \emph{ab initio} methods as the field enhancement depends on the damping and, in ultra-small quantum systems, the damping is not in general known whereas in bulk systems typically an experimental value is taken. This is particularly a problem when calculating the contributions to the Raman signal in surface-enhanced Raman scattering (SERS) for molecules close to metal clusters where both the field enhancement and chemical interactions (charge transfer between the molecule and metal) depend strongly on the linewidths of excited states \cite{Jensen2008}. Therefore, to analyse the atomic chain in terms of a plasmonic component we need an estimate of how phonons couple to electronic excitations. In a separate calculation we have used density-functional perturbation theory (DFPT) \cite{Baroni} to calculate the normal modes for sodium chains and use the highest energy mode as an estimate of the upper bound for the line-broadening of the molecular excitations. We find that high field enhancement values, comparable to what is achievable in much larger plasmonic systems, are possible in atomic chains opening up the possibility of designing quantum antennas.\\
\indent Another commonly explored plasmonic system is the dimer system which, for instance, has been explored for metallic nanospheres \cite{Hao2004}, nanorings \cite{Tsai2012} and nanorods \cite{Jain2006, Biagioni2012}. For separation distances above about $1 \ nm$ the results can be understood classically using the plasmon hybridisation model \cite{Nordlander2004}. The interference of the two LSPs leads to an extreme field enhancement and concentration in the gap region. For gap sizes below $1 \ nm$ non-local effects start to play a role and limits the achievable field enhancement \cite{Ciraci2012}. For even smaller gap sizes the electronic wavefunctions begin to overlap and electrons can tunnel across the junction forming a conductive bridge between the two MNPs and a new mode called the charge transfer plasmon (CTP) can appear \cite{Zuloaga2009}, full quantum models become essential to explain this phenomenon \cite{Zhu2016}. There have been a number of state-of-the-art experiments exploring nanogap dimer systems \cite{Ciraci2012,Savage2012,Scholl2013,Scholl2015} which have been considered in the context of SERS \cite{Zhu2014}, third harmonic generation \cite{Hajisalem2014}, strong coupling \cite{Chikkaraddy2016}, optical rectification \cite{Ward2010} and molecular tunnel junctions \cite{Tan2014}. Remarkably, given the intense interest in this area of research, there has been no excited state structural analysis of gap plasmons in ultra-small systems that we are aware of. To rectify this, we have analysed the case of a sodium chain with a gap opened in the middle and have explored how this modifies the electronic excitations and consequent plasmonic behaviour. We find that the CTPs in atomic chains can be understood as combination of a few electron-hole pairs around the HOMO-LUMO and are not very plasmon-like. We also find, analogous to larger systems, bonding dipole plasmons (BDPs) which correspond to large field enhancements in the gap region. Such knowledge may aid understanding of tunnelling in larger systems which would be difficult to model using \emph{ab initio} methods due to the large number of electrons. There are also applications in picocavity physics where the small gap region and high quality factor could create strongly mixed light-matter states \cite{Chikkaraddy2016}.

\section{Results and discussions}
\subsection{The single sodium chain}
To investigate plasmons in small molecular systems we have studied, what has become a prototypical plasmonic system to study with TDDFT, the sodium chain   \cite{Gao2005,Yan2007,Yan2008,Yasuike2011,Bernadotte2013}. We now discuss the results for a single sodium chain of 10 atoms with an interatomic distance of $d=3.08 \ \angstrom$ and aligned along the $x$ axis. The calculation techniques are discussed in appendix \ref{sec:Methods} and in appendix \ref{sec:collectivity} we show how to measure the collectivity. In figure \ref{fig:figure_1}a we present the results showing the dipole strength function. The red line shows the full Casida spectra and the blue dashed line, for comparison, shows the spectra with interactions switched off (excitations are given by the difference of Kohn-Sham (KS) orbitals). From the transition densities (density plots in figure \ref{fig:figure_1}a) one can clearly identify the longitudinal (L), transverse end (TE) and transverse central (TC) modes that were first identified by Yan \emph{et al} \cite{Yan2007} and which have also been studied for other metal chains such as gold \cite{Piccini2013}. Similar results are also seen for graphene nanoribbons \cite{Cocchi2012}. Both the L and TC modes have a dipolar character, along the long (x direction) and short lengths (y and z direction) of the chain respectively. For the widths of the peaks we use a broadening of $\eta = 0.0228 \ eV$ which is obtained by considering phonon-electron coupling and is discussed in the next section. The area under the curve must give the number of electrons according to the f-sum rule, we find that considering the spectrum up to $6 \ eV$ as well as $105$ unoccupied states satisfies the sum rule to over $96 \% $ which confirms the quality of the basis set truncation. The transverse modes are naturally higher in energy due to the greater electron confinement. The L mode is the lowest optically active mode, as is found for a classical nanorod \cite{Bryant2008}. As chain length is increased the L and TC mode grow in size whilst the TE mode saturates (this mode seems to be a consequence of the atomic nature of the ends of the chain, we have found it does not appear in simplified models where the atomic structure is ignored. In fact, it can be identified as an end mode), this accumulation in oscillator strength does not necessarily mean an increase in electron collectivity. To explore the nature of these excitations we analyse the Casida eigenvectors to see which transitions between KS orbitals make up each excitation. We find that the L mode is predominately made up of the HOMO-LUMO gap transition (states $5\rightarrow 6$ in this case). We have checked this by repeating the Casida calculation but freezing all other orbitals apart from states $5$ and $6$ which recreates the L excitation (apart from a negligible depolarization shift).\\
\begin{figure}
\includegraphics[width=8.6 cm]{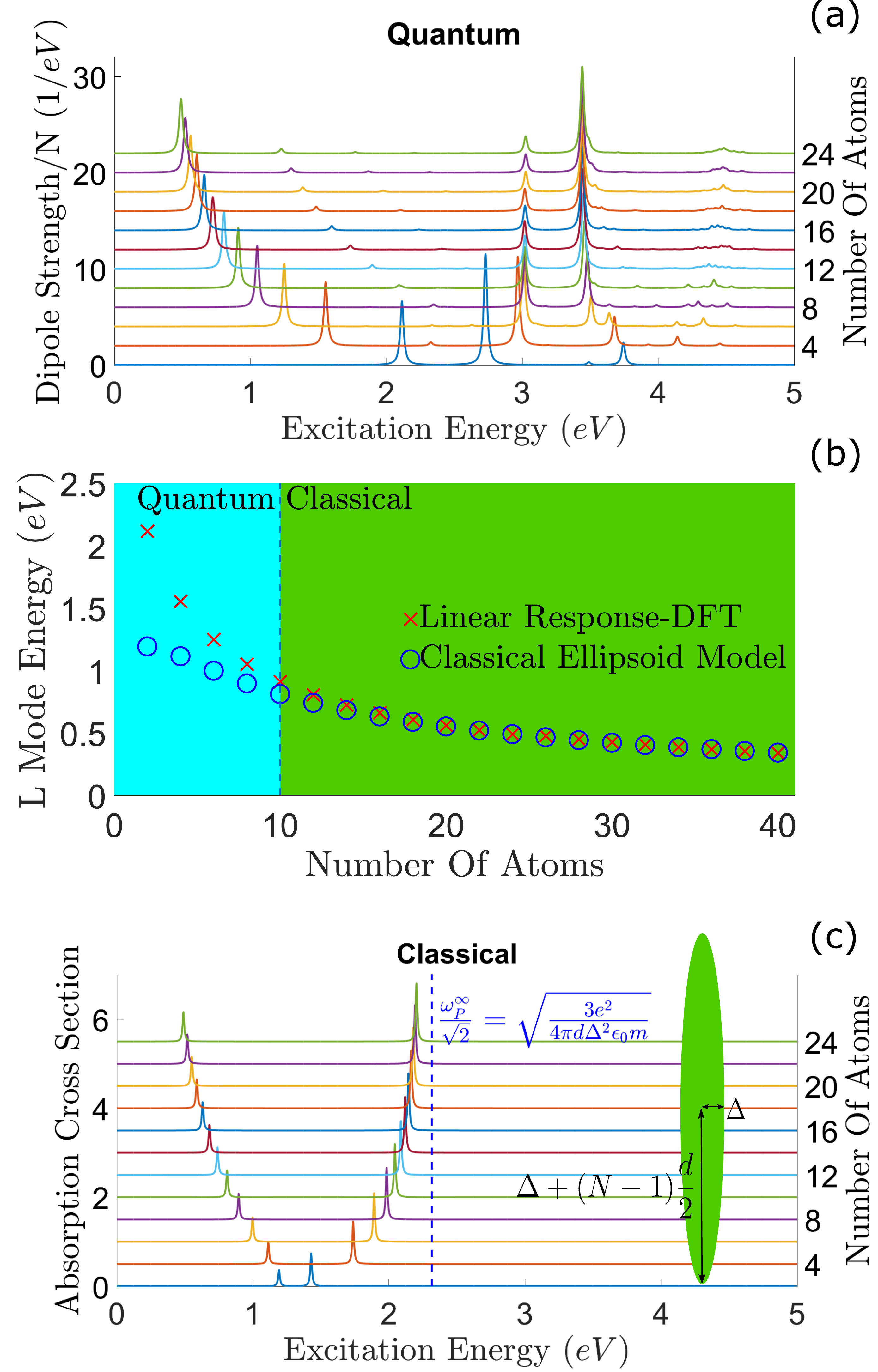}
\caption{(a) The dipole strength function (divided by the number of valence electrons) for sodium chain lengths ranging from 2 to 24 atoms. (b) Fit for the longitudinal plasmon mode using a classical ellipsoid model. The best fit for the longest chain size of $N=40$ is given by $\Delta=1.45 d$. (c) The normalised absorption cross section (averaged over 3 dimensions) calculated using the classical ellipsoid model. The dashed blue line indicates the transverse plasmon frequency for an infinite wire $\frac{\omega_p^\infty}{\sqrt{2}}$ where $\omega_p^\infty$ is the limit of the plasma frequency for infinite number of atoms.}
\label{fig:figure_2}
\end{figure}
\indent The TC excitation has a much more complicated structure than the L mode and is made up of multiple KS orbital transitions. This already hints that the TC mode is much more collective and is confirmed by the collectivity index of $4.6$: this indicates that virtually all the electrons take part in this excitation. The explanation for this collectivity is the large number of transitions with a change in the quantum number associated with the y and z direction. These have the same symmetry and are close in energy (because the length is much longer than the radius, states with different quantum numbers associated in the x direction are close in energy) which has been identified as a condition for single-particle excitations to couple \cite{Guidez2014}. This effect is increased by a degeneracy of two associated with the equivalence of the $y$ and $z$ direction. In figure \ref{fig:figure_1}b we illustrate the different KS transitions that make up the L and TC excitations \footnote{Note that there are contributions from higher energy transitions to unbound states for the TC excitation not shown in figure \ref{fig:figure_1}b.}.\\
\indent We find, in agreement with Yasuike \emph{et al} \cite{Yasuike2011}, that the L mode has a collectivity index of $1$ which led them to conclude that the L mode is a long-range charge transfer excitation rather than a plasmon. In fact, the L mode is very similar to high absorption excitations seen in conjugated polymers with high persistence lengths \cite{Vezie2016}. What complicates matters is that the L mode has been identified as plasmonic in previous works because of it scaling with the Coulomb strength, \cite{Bernadotte2013} and a large plasmonicity index \cite{Bursi2016}, furthermore the L mode is the second lowest energy excitation (rather than first which is a dark mode and so can't be seen in figure \ref{fig:figure_1}a) which hints that the Coulomb interaction does play a role but is not able to couple different transitions \cite{Piccini2013}. Thus the L mode should be identified as a `protoplasmon' - a special type of mode, peculiar to the extreme geometry of the atomic chain, that exhibits a duality of plasmonic and low collectivity behaviour. Also shown in figure \ref{fig:figure_1}a is our plasmon index indicated by the vertical black lines. For convenience we divide by the number of electrons multiplied by $\pi \eta$ (which comes from the normalisation of the Lorentzian which is convolved with the oscillator strengths to obtain a continuous spectra). We find it identifies the TC mode as the most plasmon-like excitation due to its combined dipole strength and collectivity. The TE and L mode are identified as moderately plasmonic. Interestingly the second order L mode at $2.09 \ eV$ has a much larger collectivity of $3.72$ although it is not identified as very plasmonic due to a small dipole moment. \\
\indent In figure \ref{fig:figure_2}a we show the results of a sweep over different chain lengths, from 2 to 24 atoms in steps of 2 so all chains have an even number of valence electrons (so the system is always spin unpolarized). How the excitation energy depends on the chain length can help aid identification. The L mode redshifts with increasing length which can be understood either via a decrease in the restoring force on the plasmon mode or via a decrease in the HOMO-LUMO gap. The TE and TC modes, once formed for chain lengths above about 6 atoms, do not shift as the chain width is kept constant. We find that the L mode does not become more collective as the chain size is increased. The largest chain we checked was a \ce{Na60} chain and again found a collectivity of 1 for the L mode. Interestingly it is now the 5th lowest energy excitation again indicating the role of the Coulomb interaction. The low energy absorption spectrum is shown in appendix \ref{sec:Na60_chain} revealing many higher order longitudinal modes which have increasingly higher collectivities.\\
\indent To further understand the nature of the excitations we try and identify them with classical modes, we fit the excitation energies with the two plasmon modes of a prolate spheroid which can be expressed analytically (see supplementary material). We treat the plasma frequency as dependent on the minor semiaxis $\Delta$ via the electron concentration, this reduces the number of fitting parameters and gives a more realistic test of the classical model \footnote{Similar analysis was performed in reference \cite{Yan2008} they found a surprisingly good agreement with the classical ellipsoid model by fitting both the shorter semiaxis (which will depend on the amount of electron spill-out and we will label as $\Delta$) and using the bulk plasma frequency of $3.83 \ eV$. In fact, whilst the fitting to the L mode is indeed good they seem to have fitted to the TE rather than the TC mode.}. We calculate the L mode energy for chains up to 40 atoms and for the largest chain size (which can be expected to behave the most classical) we fit for $\Delta$ and get a value of $1.45 \times d$ (see figure \ref{fig:figure_2}b). For larger chains the classical model works well but below about 10 atoms the model starts to fail and predicts energies too low which can be attributed to quantum effects as well as a failure of approximating the system as an ellipsoid. Interestingly for the above mentioned fitting parameter the TC mode energies predicted are much too low (about an $eV$) highlighting that the classical model breaks down for such small length scales, the increased energy is probably down to the strong quantum confinement. Also the classical model predicts the wrong behaviour for the (small) shift in energy of the TC mode with chain length, the classical model predicts a slight blueshift with increasing chain length for small chain sizes before becoming constant whilst the quantum calculations show a redshift. In figure \ref{fig:figure_2}c we show the corresponding, classically calculated, absorption cross section (we show only up to \ce{Na24}). We find that the relative cross section amplitudes in the classical calculation are similar to the Casida calculation showing that the identification of the TC mode with the classical transverse mode is correct and they do show some qualitative similarities. The radius of the corresponding classical ellipsoid is quite large and the electron density is practically zero at this point, this reveals why the plasma frequency is found to be much smaller than the 3D bulk value. We have also tried other methods for fitting which are shown in appendix \ref{sec:Classical_fitting} .

\begin{figure}
\includegraphics[width=8.6 cm]{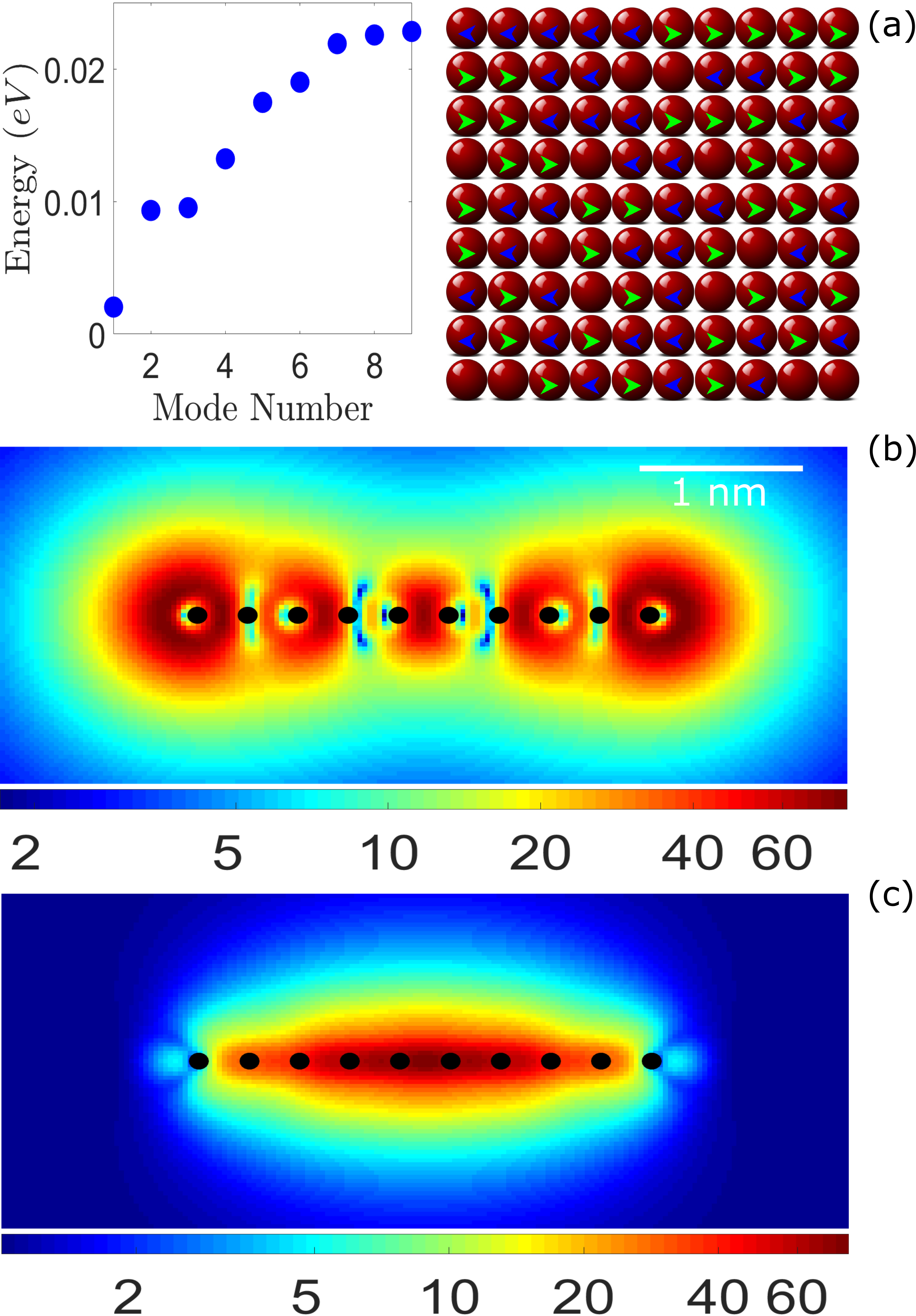}
\caption{(a) The phonon normal mode energies and corresponding atomic motion of a \ce{Na10} chain. (b) Field enhancement, calculated using equation \ref{eq:FE}, in $\log_{10}$ scale, of the L mode. (c) And TC mode. The black dots represent the sodium atom positions and the white bar indicates a length $1 \ nm$.}\label{fig:figure_3}
\end{figure}

\begin{figure*}
\includegraphics[width=17.2cm]{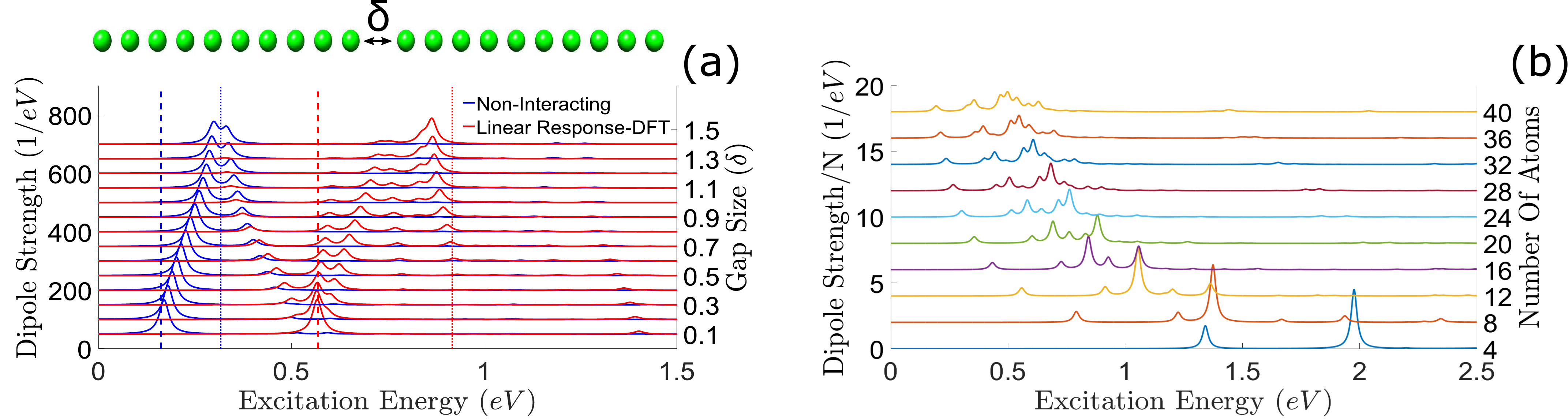}
\caption{(a) The dipole strength function of a \ce{Na20} chain with a varying gap size in the center. The gap $\delta$ is changed from $0$ to $1.4 d$, a gap size of $1d$ corresponds to a missing atom in the center. The dashed and dotted blue lines show the  \ce{Na20} and \ce{Na10} L mode energies for the non-interacting case and the red lines for the interacting. (b) Casida response for varying chain sizes from 4 to 40 atoms with a constant gap in the middle of $1 d$.} \label{fig:figure_4}
\end{figure*}

\subsection{Field enhancements}
The local field enhancement of an excitation, as induced by an electric field $\bm{E}_0 e^{-i\omega_n t}$, can be found from the induced density $\delta n$ (not to be confused with the transition density $\delta \rho$ - although they are closely connected) which can be obtained from the real part of the density matrix \cite{Casida1995} and, close to an excitation, written in the following revealing form for the Ith excitation
\begin{equation}
\delta n^I(\bm{r}, \omega_I) = - \frac{\delta \rho^I(\bm{r})}{i\eta}  \bm{\mu}^I\cdot \bm{E}. \label{eq:dn}
\end{equation}
The response is purely imaginary as the electrons oscillate out of phase at the resonance. In appendix \ref{sec:field_enh} we show a derivation of this equation, it has also been derived, in a different manner, by Cocchi \emph{et al} \cite{Cocchi2012}. The field enhancement
\begin{equation}
FE = \frac{|\delta \bm{E}(\bm{r},\omega)+\bm{E}_{ext}|}{|\bm{E}_{ext}|} \label{eq:FE}
\end{equation}
can be obtained within the electrostatic approximation. Clearly the induced field depends crucially on the magnitude of the damping for the excitation. Using DFPT we calculate the vibrational normal modes of the \ce{Na10} chain. There are a total of $3N=30$ modes with only the top $9$ modes corresponding to non-zero energies. In figure \ref{fig:figure_3}a we show the excitation energies and the corresponding motion of the sodium atoms which all turn out to correspond to longitudinal motion. We take the largest energy of $0.0228 \ eV$ and use it as the linewidth in the Casida calculations. This is the simplest method to include electron-phonon coupling and should give an upper-bound on the true value.
 We note that this is a particularly simple method and assumes the phonon is a perturbation on the electronic spectra. For larger plasmon-phonon coupling there will be richer spectral features as recently shown for aromatic hydrocarbons \cite{Cui2016}. For longer chains the L mode energy reduces and the phonon energy increases as shown in figure \ref{fig:figure_3}a which suggest for certain chain lengths there should be a strong-coupling regime. In figures \ref{fig:figure_3}b and \ref{fig:figure_3}c the field enhancement in $\log_{10}$ scale is shown for the L and TC mode respectively. We find the maximum field enhancement to be $80.5$ for the L mode and $78.3$ for the TC mode (which corresponds to intensity enhancements of $6400$), these values are comparable to the field enhancements in larger nanostructures. The electric field is qualitatively very similar to that seen for a classical antenna and shows that the modes will couple to the far field. From a practical point of view, whilst the maximum field enhancements are attractive, the sharp fall off of the induced field, due to the small number of electrons, may limit potential plasmonic applications. Fortunately this does lead to extremely large field gradients that could be useful in exciting higher order multipole excitations in small molecules and atoms, for example the TC mode has a maximum gradient enhancement $\nabla |\bm{\delta E}|/|\bm{E}_{ext}|$ of $17.1 \ a_0^{-1}$ along the z direction for the slice $x=y=0$. If one makes do with more modest field enhancement then molecular antennas may be useful for inducing large local fields: for the TC mode field enhancements of around 5 are achievable at about $2.2$ bond lengths away which hints it might be possible to alter the optical density of states with minimal electronic coupling between the chain and a small absorbing molecule.

\subsection{Double sodium chains}
Next we consider separating an atomic chain along the longitudinal direction. First we consider taking a \ce{Na20} chain and separating it in the middle, figure \ref{fig:figure_4}a shows the low energy spectra as the gap length is changed as we separate the gap ($\delta$) up to $1.4 d$, we resist going further as the LDA is not suitable to consider long range van der Waals interactions which will dominate. We find that as the gap is opened there is generation of a number of new modes and we find a qualitatively similar mode to the low energy charge-transfer plasmons that has been explored heavily in the context of nearly touching dimers \cite{Zhu2016}. The non-interacting electronic spectra can be understood as a splitting of energy levels due to the gap and the creation of symmetric and anti-symmetric pairs of eigenstates. The transverse modes (for clarity not shown in figure \ref{fig:figure_4}a, the full spectra up to $4 \ eV$ is shown in appendix \ref{sec:Transverse_gap_plasmons}) are relatively unchanged via the presence of a gap as is the case for classical nanorod dimers \cite{Jain2006}, we do though observe some interference effects between the TC and TE modes at small gap sizes. There is also the formation of a number of new longitudinal modes for gap sizes $0.4d \rightarrow 1.4 d$, which we will label as `tunnelling' or `gap' modes as they involve the transfer of charge across the gap, due to electron tunnelling, every optical cycle. We find that they tend to have a low collectivity compared to the transverse modes, but crucially they have a larger collectivity than the L mode of a single chain indicating that they arise as a consequence of multiple electron-hole excitations. For larger separation distances ($>1.2 d$) the low energy absorption is dominated by the longitudinal BDP which originates from the electrostatic coupling of the two chains and is red-shifted from the mode that would be present with no coupling. To identify the peaks from the complicated spectra we will take a practical definition of a CTP as involving each chain having a different sign of charge and the intensity should go to zero for large gap sizes. The BDP should have a dipole charge distribution for each chain and the intensity will not go to zero for large distances but rather merge with the L mode of a single \ce{Na10} chain. Identification is easiest using the induced potential (see appendix \ref{sec:ind_pot_gap_plasmons}). For small gap sizes, where electron tunnelling is most important, two CTPs dominate the spectra. The lowest energy excitation we label CTP1 which redshifts and loses intensity as the gap increases and becomes negligible above a gap size of about $1 d$. There is also another CTP which we label CTP2 which also loses intensity as the gap sizes increase. Both CTP modes merge with the \ce{Na20} L mode as $\delta \rightarrow 0$. We identify a single BDP that starts to become apparent about $0.7 d$ and grows in intensity, it is the highest energy tunnelling mode and is close in energy to the \ce{Na10} L mode. In between the CTPs and the BDP there are 3 modes which we believe are best explained as hybrid modes: they have a similar charge distribution as a BDP although the change in the potential over the gap region is smaller which is a consequence of charge transfer neutralising the induced charge setup across the gap, they also die out as the gap length is increased. This is a pleasing example of how plasmonic concepts and analogies can be used to simplify analysis of an electronic excitation spectra, as compared to analysis of the orbitals, which in this case is complicated and not very illuminating.\\
\begin{figure}
\includegraphics[width=8.6 cm]{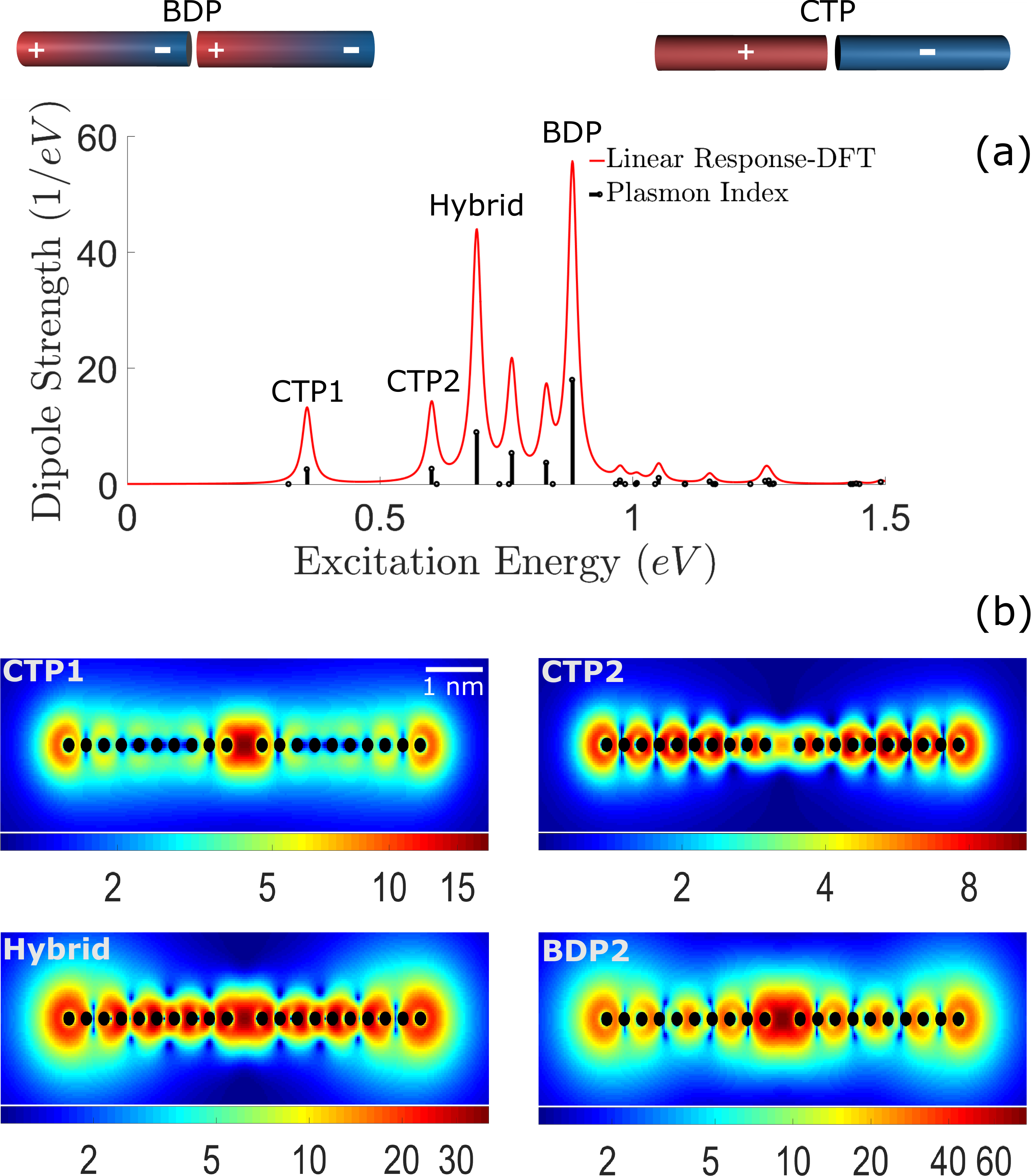}
\caption{(a) The dipole strength function of \ce{Na20} with a gap of $1 d$ (i.e. one atom removed) in the center. (b) The field enhancements in $\log_{10}$ scale of the CTP1, CTP2, hybrid and BDP modes.} \label{fig:figure_5}
\end{figure}
\indent To aid identification of the modes we have also considered holding the gap distance fixed and sweeping the chain length: see figure \ref{fig:figure_4}b. We consider only the lower energy L modes allowing us to consider chain sizes up to 40 atoms, we only look at atomic numbers that are multiples of 4 to make sure that the separated chains have an even number of valence electrons each to ensure the validity of unpolarized spin calculations. We find that the CTPs are present for very small chains highlighting that they do not involve many electrons and originate from the interaction of electronic wavefunctions near the HOMO-LUMO gap. Crucially the L modes do not become more plasmonic as the chain length is increased.\\
\indent Next we look in detail at the case of a gap size of $1d$ (which is equivalent to removing an atom from the middle of a 21 atom chain) as shown in figure \ref{fig:figure_5} where the low energy spectra and the field enhancement of 4 of the tunnelling modes is shown. We find that at this distance our plasmon index predicts the hybrid mode and BDP to be much more plasmonic excitations than CTP1 and CTP2, this is backed up by the calculated field enhancement which reaches maximum values of $38.2$ and $80.1$ for hybrid and BDP modes respectively compared to $18$ and $10.6$ for the CTP1 and CTP2 (see figure \ref{fig:figure_5}b). The large plasmonicity for the BDP comes from its large dipole moment rather than a large collectivity, although of the 4 modes analysed it has the largest collectivity of $3.29$ as compared to values close to $2$ for the other 3 modes. The BDP mode is also by far the most interesting in terms of potential applications as it exhibits the largest field enhancement in the gap region where potentially a small atom could be placed and its emission and absorption properties modified. Thus atomic chains could potentially be used as a picocavity and modify light-matter interactions on the smallest possible scale.

\section{Conclusions}
In conclusion we have used TDDFT to study the optical response of single-atom-thick atomic chains. We have identified plasmon modes using a new measure of `plasmonicity' based on excited state structural analysis and have found that there are two modes that correspond to the classical plasmon modes but the transverse mode is blueshifted because of the extreme thinness. For the \ce{Na10} chain we have used DFPT, in a separate calculation, to calculate the phonon modes and have used the highest energy normal mode to estimate the upper bound of the linewidth and hence calculate the field enhancement to test the suitability of the atomic chain as a quantum antenna. We found that high field enhancements are achievable and the rapid spatial decay of the field away from the molecule, while it may limit some potential application, results in large field gradients. Finally, we have considered an atomic chain dimer system and have explored the spectra and excited-state structure of low energy tunnelling modes as a function of gap size and chain size. We find that they are not very collective but do bear some similarities to the charge transfer plasmons of larger dimer systems in their spectral behaviour and charge distribution. A large field enhancement can be achieved in the gap region in analogy to dimers in classical plasmonics. This work shows there is pleasing synergy between the ideas of plasmonics and electronic structure which will aid development of new quantum nano-optic devices.

\begin{acknowledgments}
The work of J. M. Fitzgerald was supported
under a studentship from the Imperial College London funded by the EPSRC
Grant 1580548.
\end{acknowledgments}

\appendix
\section{Methods}\label{sec:Methods}
We used the open source real-space code OCTOPUS \cite{Andrade2015}. The sodium atoms are described by a norm-conserving Troullier-Martins pseudopotential \cite{Troullier1991} with only the 3s electrons explicitly modelled in the calculation. The (modified) local density approximation (LDA) \cite{Perdew1981} was used for the exchange-correlation potential for both the ground-state and for excited-state calculations, this is known to work well for sodium. We use the simplest exchange-correlation functional as it is well known that the existence of plasmons depends only on the Coulomb interaction and that exchange-correlation effects only provide a small correction (a small red shift) to the plasmon energy. All fields, such as the KS wavefunctions and the density, are represented on equally spaced real-space grids, the simulation box is built via a union of spheres centred on each atom. We used a radius of $12 \ \angstrom$ and a grid spacing of $0.3 \ \angstrom$ which gives a good compromise between accuracy and speed. For the spacing of the sodium atoms we use the interatomic distance of \ce{Na2} dimers of $d = 3.08 \ \angstrom$ \cite{Vasiliev2002}, no attempt was made of structural optimisation of any of the structures.\\
\indent The electronic spectra is calculated using the Casida method \cite{Casida1995,Jamorski1996} which is a representation of linear response theory in a configuration (particle-hole) space and is similar in spirit to the matrix formulation of RPA \cite{Bertsch1994}. It involves solving a (pseudo) eigenvalue equation to find the electronic eigenmodes of the system
\begin{equation}
\begin{pmatrix}
\bm{A} & \bm{B} \\
\bm{B}^* & \bm{A}^*
\end{pmatrix}
\begin{pmatrix}
\bm{X}\\
\bm{Y}
\end{pmatrix}
= \omega
\begin{pmatrix}
\bm{X}\\
\bm{Y}
\end{pmatrix} \label{eq:Casida}
\end{equation}
where the elements of the matrices are
\begin{equation}
\begin{split}
A_{ia,jb} = (\epsilon_a-\epsilon_i)\delta_{ij}\delta_{ab} + 2 K_{ia,jb}\\
B_{ia,jb} = 2 K_{ia,bj}
\end{split}
\end{equation}
and $K$ is the two-body Coulomb and exchange-correlation integral constructed with the KS eigenstates within the adiabatic approximation
\begin{equation}
\begin{split}
K_{ia,jb} = \int d\bm{r} d\bm{r}' \  \phi_i^*(\bm{r})\phi_a(\bm{r}) \\
 \times \left(\frac{1}{|\bm{r}-\bm{r}'|} + f_{XC}(\bm{r},\bm{r}') \right) \phi_b^*(\bm{r}') \phi_j(\bm{r}') 
\end{split}
\end{equation}
(we use the standard convention that $i,j$ and $a,b$ represent occupied and unoccupied orbitals respectively. The number of occupied and unoccupied orbitals will be given by $N_o$ and $N_u$). The $N_oN_u$ eigenvalues give the excitation energies and the eigenvectors can calculate physically relevant quantities like the transition density $\delta \rho$, dipole moments $\delta \bm{\mu}$ and dipole strength function.\\
\indent It is well known that the one-particle transition density matrix $T_{ia}$ encodes, in a convenient and compact fashion, information on electronic transitions from the ground state to the excited states. It is in general non-symmetric so cannot be diagonalised, instead one can perform a singular value decomposition to express it in natural transition orbitals (NTOs) in which it will be diagonal \cite{Martin2003}
\begin{equation}
[ \bm{U}^\dagger \bm{T} \bm{V}]_{ij} = \sqrt{\lambda_{i}}\delta_{ij},
\end{equation}
where $\bm{U}$ and $\bm{V}$ are unitary transformations from the ground state orbitals to the NTOs for the occupied and virtual orbitals respectively. The eigenvalue $\lambda_i$ represents the contribution of the ith electron-hole pair to a particular excitation and has the normalisation condition $ \sum_i \lambda_i = 1$. In the context of plasmonics it is useful to define a collectivity index (also called the inverse participation ratio) \cite{Martin2003,Plasser2014}
\begin{equation}
R = \frac{1}{\sum_i \lambda_i^2}
\end{equation}
The collectivity index of an excitation measures the number of electron-hole pairs that contribute: the higher the value the more collective the excitation. If there is no electronic correlation, then each excitation represents a single transition and the index is equal to 1. Thus we have a quantitative measure of an excitations collectivity which will clearly be useful in identifying which excitations are plasmon-like. But on its own it is not a useful measure as many collective excitations are dipole forbidden and will not be excited by light, a useful plasmonic mode will couple strongly to light. With this in mind we introduce a new plasmon index given by the product of the collectivity and the dipole strength (of course one is not always only interested in dipole modes, for instance quadrupole modes could be of interest - in which case our index could be adjusted accordingly).\\
\indent One can calculate the KS transition density matrix from a Casida calculation allowing the collectivity of an excitation to be calculated. Any excitation can be written as a superposition of p-h states with the eigenvectors of equation \ref{eq:Casida} acting as coefficients describing the contribution of each possible electron-hole transition. This clearly suggests that the Casida eigenvectors must contain information on the collectivity of an excitation. For practical considerations the Casida calculations are not solved in the form of equation \ref{eq:Casida} and direct access to the vectors $X$ and $Y$ is not possible in OCTOPUS. Also the presence of de-excitations (described by the $\bm{Y}$ eigenvector) complicates calculating the transition density matrix (within the Tamm-Dancoff approximation, where there is no coupling between the hole-particle and hole-particle correlations, it is possible to simply identify that $T_{ia} = X_{ia}$ \cite{Etienne2015}). In appendix \ref{sec:collectivity} the necessary post-analysis of typical TDDFT results is shown.\\
\indent Phonon spectra and normal modes are obtained using DFPT \cite{Baroni} as implemented in Quantum Espresso code \cite{QE}. We extensively employed DFPT for calculation of lattice dynamics and phonon energies of crystals \cite{PRB13,PRB14,PRL14} and molecules \cite{JCP16}. As for the TDDFT calculations, we used the Perdew-Zunger \cite{Perdew1981} parametrization of the local-density-approximation (LDA) for the exchange-correlation functional. In our density functional calculations, a non-relativistic norm-conserving pseudopotential is used on an which 3s orbital with an occupation of 1 is determined as the non-local channel. Electronic structure results are calculated using the energy cutoff $60 \ Ry$. For calculations on isolated one-dimensional Na-chain, we employ open boundary conditions. We performed this by locating the Na-chain in a large simulation box and truncating the Coulomb interaction at long distances to eliminate electrostatic interactions between periodic images.

\section{Calculation of the collectivity}\label{sec:collectivity}
Typically a slight variation (valid for real wavefunctions) of the Casida equation (equation \ref{eq:Casida}) is solved in electronic structure codes
\begin{equation}
\bm{C}^+\bm{Z}^+= \omega^2 \bm{Z}^+
\end{equation}
where
\begin{equation}
\begin{split}
\bm{C}^+ = (\bm{A}-\bm{B})^{\frac{1}{2}}(\bm{A}+\bm{B})(\bm{A}-\bm{B})^{\frac{1}{2}}\\
\bm{Z}^+ = \sqrt{\omega}(\bm{A}-\bm{B})^{-\frac{1}{2}}(\bm{X}+\bm{Y})
\end{split}
\end{equation}
and $\bm{X}$, $\bm{Y}$, $\bm{A}$ and $\bm{B}$ were defined in the main text. This is a convenient transformation as it reduces the dimensionality of the problem by a half and one needs only deal with one set of eigenvectors when calculating physical quantities such as the transition density and dipole oscillator strength. It is also now an Hermitian eigenvalue problem. What is often overlooked is that there is an alternative and equivalent form of these equations \cite{Luzanov2010}
\begin{equation}
\bm{C}^-\bm{Z}^-= \omega^2 \bm{Z}^-
\end{equation}
where
\begin{equation}
\begin{split}
\bm{C}^- = (\bm{A}+\bm{B})^{\frac{1}{2}}(\bm{A}-\bm{B})(\bm{A}+\bm{B})^{\frac{1}{2}}\\
\bm{Z}^- =  \sqrt{\omega}(\bm{A}+\bm{B})^{-\frac{1}{2}}(\bm{X}-\bm{Y}).
\end{split}
\end{equation}
Both eigenvectors are normalised to one 
\begin{equation}
{|\bm{Z}^{\pm}|}^2 = 1
\end{equation}
which is consequence of the more unusual non-euclidean normalisation condition of the Casida eigenvectors
\begin{equation}
|\bm{X}|^2-|\bm{Y}|^2 = 1.
\end{equation}
The two sets of eigenvectors are related by the equation
\begin{equation}
\bm{Z}^{-} = \frac{(\bm{A}+\bm{B})^{\frac{1}{2}} (\bm{A}-\bm{B})^{\frac{1}{2}}}{\omega} \bm{Z}^+
\end{equation}
where the square roots should be understood as principal square roots of matrices. Thus if one has access to the two-body matrix elements $K_{ia,jb}$ then it is simple to construct both matrices $\bm{C}^{\pm}$ and then solving a simple Hermitian eigenvalue equation to obtain $\bm{Z}^{\pm}$. It is also possible to calculate the eigenvectors $\bm{X}$ and $\bm{Y}$ using the equations
\begin{equation}
\begin{split}
\bm{X} = \bm{D}^- \bm{Z}^+ + \bm{D}^+ \bm{Z}^-\\
\bm{Y} = \bm{D}^- \bm{Z}^+ - \bm{D}^+ \bm{Z}^-
\end{split}
\end{equation}
where
\begin{equation}
\begin{split}
\bm{D}^+ = \frac{(\bm{A}+\bm{B})}{2 \sqrt{\omega}}\\
\bm{D}^- = \frac{(\bm{A}-\bm{B})}{2 \sqrt{\omega}}
\end{split}
\end{equation}
\\
\indent In our experience for the systems considered in this work, taking the transition density matrix to be either $\bm{Z}^+$ or $\bm{Z}^{-}$ provides reasonable results but the non-uniqueness is not satisfactory. There has been some work on uniquely determining the transition density matrix including \emph{ph-hp} correlation \cite{Etienne2015,Luzanov2010,Yasuike2011}. Given that the different methods return similar values we use the simplest method from \cite{Yasuike2011} where two pseudo transition matrices are defined and then we use the following equation for the collectivity
\begin{equation}
R = R_X^{(\sum_{ia}X_{ia}^2)}R_Y^{(\sum_{ia}Y_{ia}^2)}.
\end{equation}

\section{Calculation of the field enhancement}\label{sec:field_enh}
To calculate the field enhancement using the results of a Casida simulation we begin by considering the real part of the density matrix \cite{Casida1995}
\begin{equation}
\Re\left(\delta P_{ia\sigma}(\omega)\right) = \sum_{jb\tau} \frac{1}{\omega^2-C_{ia\sigma,jb\tau}}  v_{jb\tau}^{appl}(\omega)
\end{equation}
where $v_{jb\tau} = \int d^3r \phi_{j\tau}(\bm{r}) v^{appl}(\bm{r},\omega)\phi_{b\tau}(\bm{r})$ and $v^{appl}$ is the perturbation. This equation can be rewritten using, a spectral expansion, in terms of the Casida eigenvalues and eigenvectors
\begin{equation}
\begin{split}
\Re\left(\delta P_{ia\sigma}(\omega)\right)  = \sum_{I} \frac{1}{\omega^2-\omega_I^2}\\
\times \sum_{jb\tau} \sqrt{\omega_{ai\sigma}} Z_{ia\sigma}^I \left({Z_{jb\tau}}^I\right)^*\sqrt{\omega_{bj\tau}} v_{jb\tau}^{appl}(\omega).
\end{split}
\end{equation}
The induced density can be defined as \cite{Casida1995}
\begin{equation}
\delta n(\bm{r},\omega) = \sum_{ia\sigma} \phi_{i\sigma}(\bm{r})\phi_{a\sigma}(\bm{r}) \delta P_{ia\sigma}(\omega)
\end{equation}
and, if we consider a perturbation of the form $v^{appl}(\bm{r},\omega) = \bm{r} \cdot \bm{E}$, then
\begin{equation}
\begin{split}
\delta n(\bm{r},\omega) = 
\sum_{I} \frac{2}{\omega^2-\omega_I^2} \\
\times \sum_{ia\sigma,jb\tau} \phi_{i\sigma}(\bm{r})\phi_{a\sigma}(\bm{r})\sqrt{\omega_{ai\sigma}} Z_{ia\sigma}^I \left({Z_{jb\tau}}^I\right)^*\sqrt{\omega_{bj\tau}} \bm{r}_{jb\tau} \cdot \bm{E}(\omega).
\end{split}
\end{equation}
In its present form the pole structure of the equations means the induced density diverges at the excitation energies, thus we add a finite life time correction $\eta$
\begin{equation}
\begin{split}
\delta n(\bm{r},\omega) = 
\sum_{I} \frac{1}{\omega_I} 
\left(\frac{1}{\omega-\omega_I+i\eta/2} - \frac{1}{\omega+\omega_I+i\eta/2} \right)\\
 \times \sum_{ia\sigma,jb\tau} \phi_{i\sigma}(\bm{r})\phi_{a\sigma}(\bm{r})\sqrt{\omega_{ai\sigma}} Z_{ia\sigma}^I \left({Z_{jb\tau}}^I\right)^*\sqrt{\omega_{bj\tau}} \bm{r}_{jb\tau} \cdot \bm{E}(\omega).
 \end{split}
\end{equation}
which agrees with the result of Rossi \cite{Rossi2013}. Usually one is most interested in frequencies at the Casida excitation energies, therefore we can save some computational effort by expanding about the point $\omega = \omega_I$ and reordering the summations to get our final results
\begin{equation}
\delta n(\bm{r},\omega \approx \omega_I)  \approx  -\frac{1}{i\eta} \delta \rho^I(\bm{r}) \bm{\mu}^I\cdot \bm{E}. \label{eq:sup:ind_dens}
\end{equation}
Where we have used the following definitions for the transition density and transition dipole moment for the transition between the many body states $\Psi_0$ to $\Psi_I$ in the Casida formalism \cite{Casida1995,Yasuike2011}
\begin{equation}
\begin{split}
\delta \rho^I(\bm{r}) = \braket{\Psi_0|\sum^N_i \delta(\bm{r}-\bm{r}_i)|\Psi_I}\\
 = \sqrt{2} \sum_{ia} \phi_i(\bm{r}) \phi_a(\bm{r}) \left( X_{ia}^I + Y_{ia}^I \right)
 \end{split}
\end{equation}
and
\begin{equation}
\delta \bm{\mu}^I(\bm{r}) = \braket{\Psi_0|\hat{\bm{\mu}}|\Psi_I} = \sqrt{2} \sum_{ia} \bm{\hat{\mu}}_{ia} \left( X_{ia}^I + Y_{ia}^I \right)
\end{equation}
where we assume real KS orbitals and $\bm{\mu}_{ia} = - \int \  \phi_i(\bm{r}) \phi_a(\bm{r}) \ \bm{r} \ d^3 r$. Equation \ref{eq:sup:ind_dens} is an intuitive result and has been derived via a different method by Cocchi \emph{et al} \cite{Cocchi2012}. The induced density for a particular excitation has a spatial profile given by the transition density and its magnitude is given by the projection of the electric field on the transition dipole moment, which describes how strongly the excitation couples to the perturbation. It is also inversely proportional to the linewidth of the excitation. The induced density can then be used to calculate the induced electric field and the consequential field enhancement.

\section{Na60 chain}\label{sec:Na60_chain}

\begin{figure}
\includegraphics[width=8.6cm]{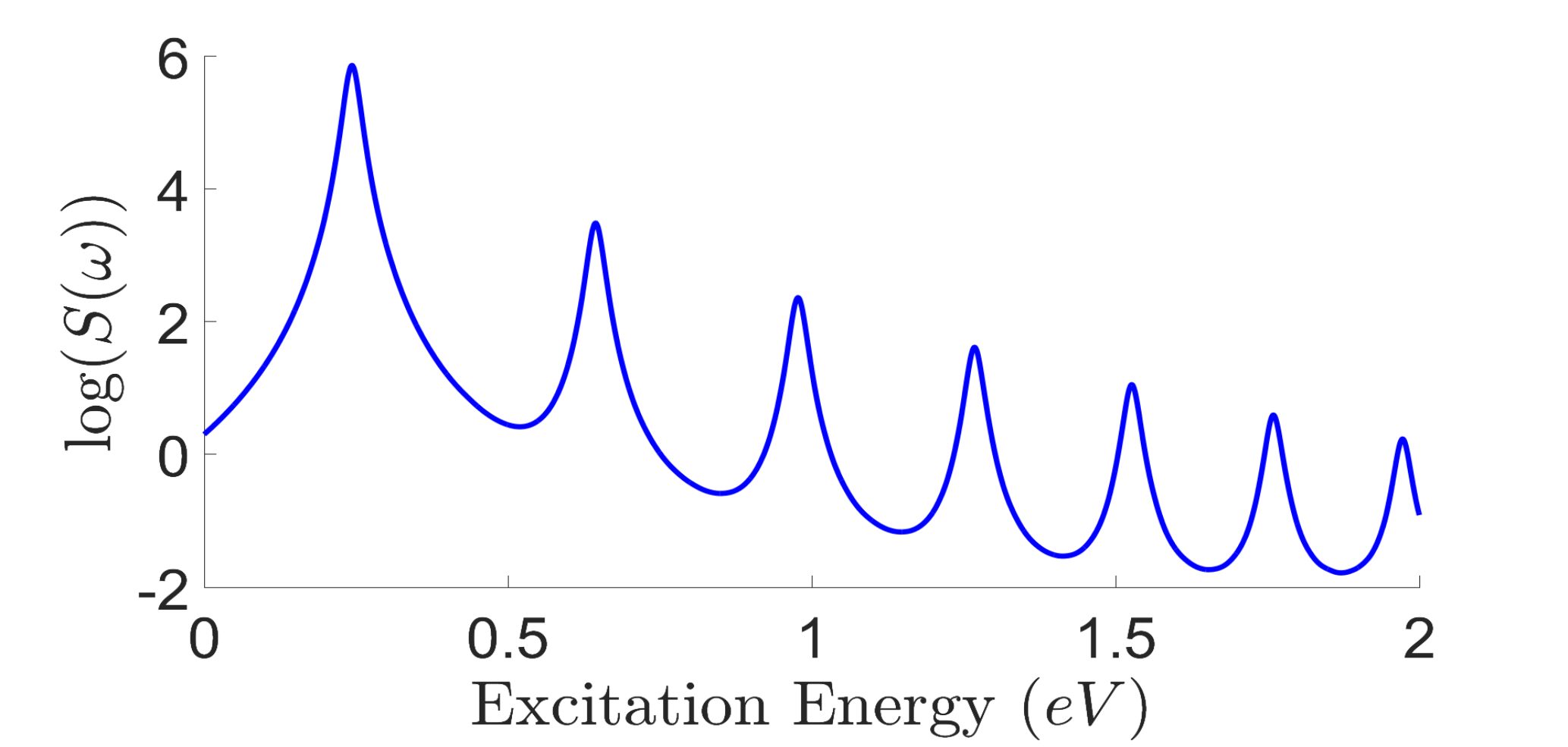}
\caption{The logarithmic dipole strength function for a \ce{Na60} chain.} \label{fig:Na60_Spectrum}
\end{figure}

In figure \ref{fig:Na60_Spectrum} results for a Sodium Na60 chain with the spectrum calculated up to $2 \ eV$ in figure \ref{fig:Na60_Spectrum}. The chain has $60$ occupied states and a total of 80 unoccupied states were considered. The log of the dipole strength function is taken to clearly show the higher order longitudinal plasmon modes. As the chain size is increased these modes bunch together and in the infinite limit give the dispersion curve of a plasmon in a 1D electron gas \cite{Bernadotte2013}.

\begin{figure*}
\includegraphics[width=17.2cm]{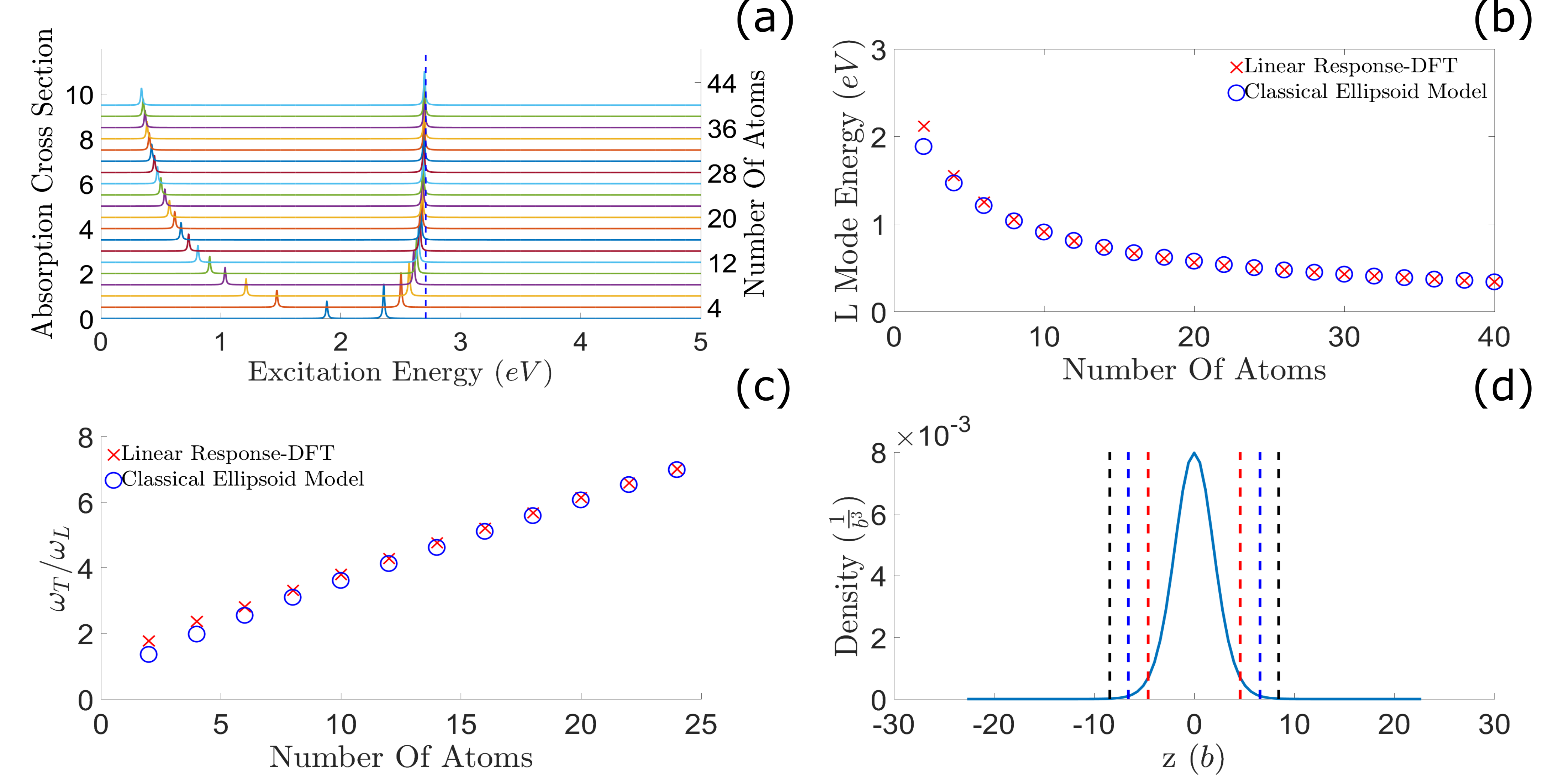}
\caption{(a) The normalised absorption cross section calculated using the classical ellipsoid model for Na chains ranging from 2 to 40. The dashed blue line indicates the transverse plasmon frequency using $\omega_p =3.83 \ eV$. (b) Fit for the longitudinal plasmon mode using a classical ellipsoid model. The best fit for the longest chain size of $N=40$ is given by $\Delta=1.13 d$. (c) The ratio of the transverse and longitudinal plasmon energy calculated from the Casida and the classical ellipsoid method. (d) Cross section of the ground state density for a \ce{Na10} chain for $x=0, y=0$ with the fitting parameters $\Delta=1.45d, 1.13d$ and $0.79d$ are indicated by the black, blue and red dashed line respectively. } \label{fig:Sup_figure_2}
\end{figure*}

\section{Fitting to the classical model}\label{sec:Classical_fitting}
Here we fit our results to a classical ellipsoid model within the electrostatic approximation \cite{Bohren2008}. The conditions for a localised resonance of an ellipsoid, with semiaxes $a,b,c$ is given by poles of the polarizability
\begin{equation}
\epsilon_m + L_i \left(\epsilon(\omega)-\epsilon_m \right) = 0
\end{equation}
where $i=\{a,b,c\}$, $\epsilon_m$ is the permittivity of the surrounding medium (we take it to be one). We are interested in the particular case of a prolate spheroid where $b=c$ and $a$ is the largest semiaxes, in this case $L_a$ can be found analytically
\begin{equation}
\begin{split}
L_a = \frac{1-e^2}{e^2}\left(\frac{1}{2e}\log{(\frac{1+e}{1-e})}-1\right)\\
e^2 = 1 - \frac{b^2}{a^2} = 1- \frac{c^2}{a^2} 
\end{split}
\end{equation}
where $e$ is the eccentricity of the ellipsoid. Following Yan and Gao \cite{Yan2008} we define $a=\Delta+(N-1)\frac{d}{2}$ and $b=c=\Delta$ where $d=3.08 \angstrom$ is the bonding length used in the simulation and $\Delta$ is a measure of the electron spill out at the edges of the chain and is used as a fitting parameter. Yan and Gao \cite{Yan2008} used a fixed bulk value for the plasma frequency in the Drude dielectric function used for $\epsilon(\omega)$, effectively adding another fitting parameter. Instead we define a chain size dependent plasma frequency
\begin{equation}
\omega_p^2 = \frac{n e^2}{\epsilon_0 m} = \frac{N e^2}{V \epsilon_0 m} = \frac{3 N e^2}{4 \pi a b c \epsilon_0 m} 
= \frac{3 N e^2}{4 \pi (\Delta+(N-1)d/2) \Delta^2 \epsilon_0 m}
\end{equation}
meaning we need only fit $\Delta$. For large $N$ the plasma frequency becomes independent of $N$ as one would expect. Once we have fitted for $\Delta$ we can then calculate $L_b$ via the relation
\begin{equation}
L_a+L_b+L_c=L_a+2L_b = 1.
\end{equation}
Applying the above procedure, we obtain the results show in the main text where $\Delta = 1.45 d$. The polarizability can be calculated using
\begin{equation}
\alpha_i = 4 \pi abc \frac{\epsilon(\omega)-\epsilon_m}{\epsilon_m + L_i \left(\epsilon(\omega)-\epsilon_m \right)}
\end{equation}
from which the (averaged over the 3 spatial dimensions) absorption cross section (which dominates over scattering for small particles) can be calculated
\begin{equation}
\sigma_{abs} = \frac{2 \pi}{3 \lambda} \Im [ \sum_i \alpha_i(\omega) ]
\end{equation}
which was used to obtain the results in figure 2c in the main text. 
\\
\indent If we instead we take the plasma frequency to be fixed at the bulk value (we use $\omega_p =3.83 \ eV$ as used by Yan and Gao \cite{Yan2008}) then we obtain a best fit of $\Delta = 1.13 d$. The absorption cross section and the results of the fit for different chain sizes are shown in figure \ref{fig:Sup_figure_2}a and b respectively.\\
\indent Another potential method is to consider only the ratio of the T and L mode energies $\frac{\omega_T}{\omega_L}$ which is convenient as the plasma frequency does not need to be considered. One can find that
\begin{equation}
\frac{\omega_T}{\omega_L} = \sqrt{\frac{1-L_a}{2 L_a}}
\end{equation}
which can be compared to the linear response DFT results. The fitting parameter is obtained for the largest chain and comes out to $\Delta =0.79 d$ and the results are shown in see figure \ref{fig:Sup_figure_2}c. In figure \ref{fig:Sup_figure_2}d we show a cross section view of the ground state density, along the z direction, calculated by DFT and the radii for the various fits discussed are shown in comparison.

\begin{figure}
\includegraphics[width=8.6cm]{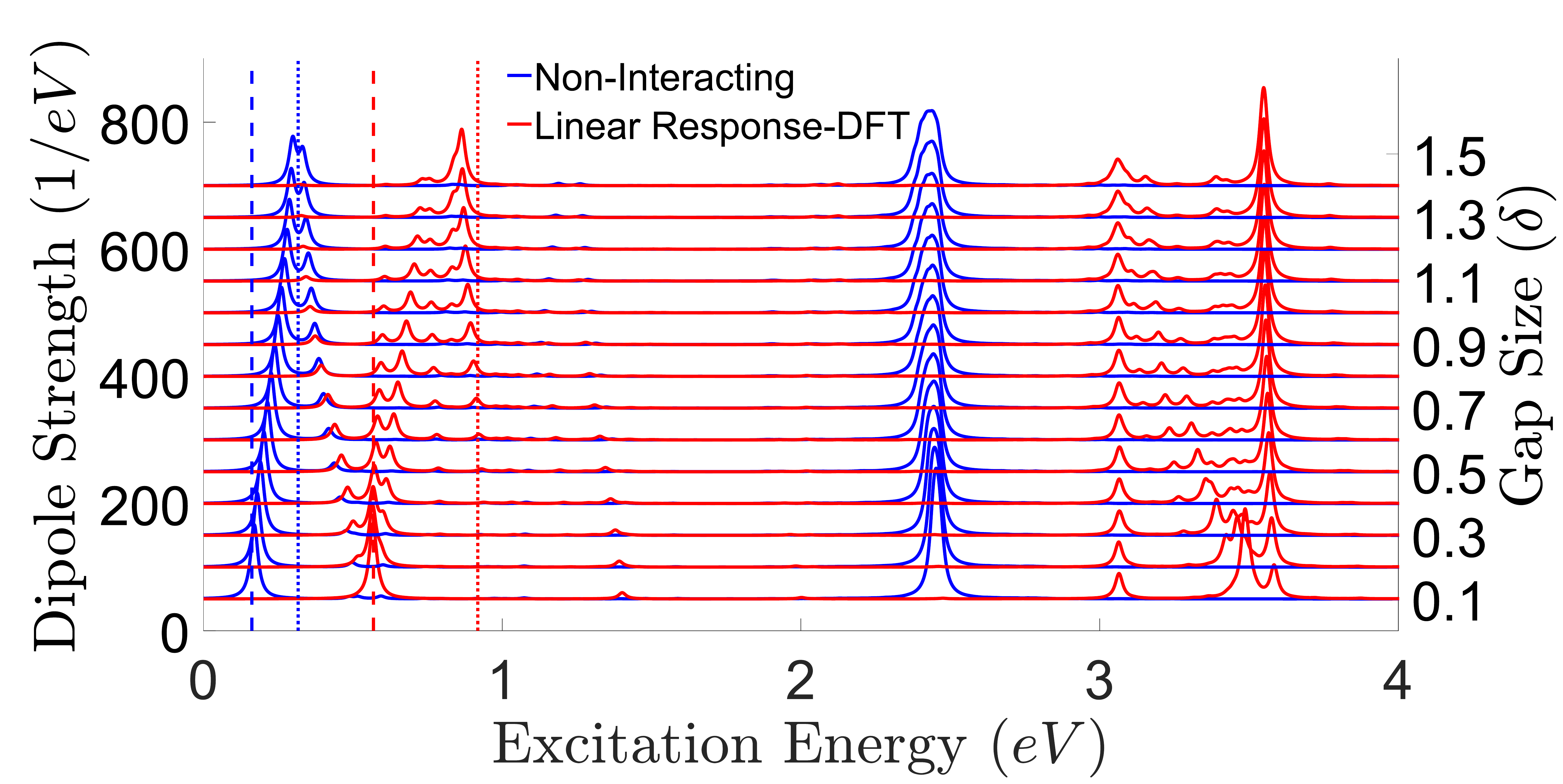}
\caption{The dipole strength function of a \ce{Na20} chain with a varying gap size in the center. The gap is changed from $0$ to $1.4 d$, a gap size of $1$ corresponds to a missing atom in the center.} \label{Sup_figure_4}
\end{figure}

\begin{figure}
\includegraphics[width=8.6cm]{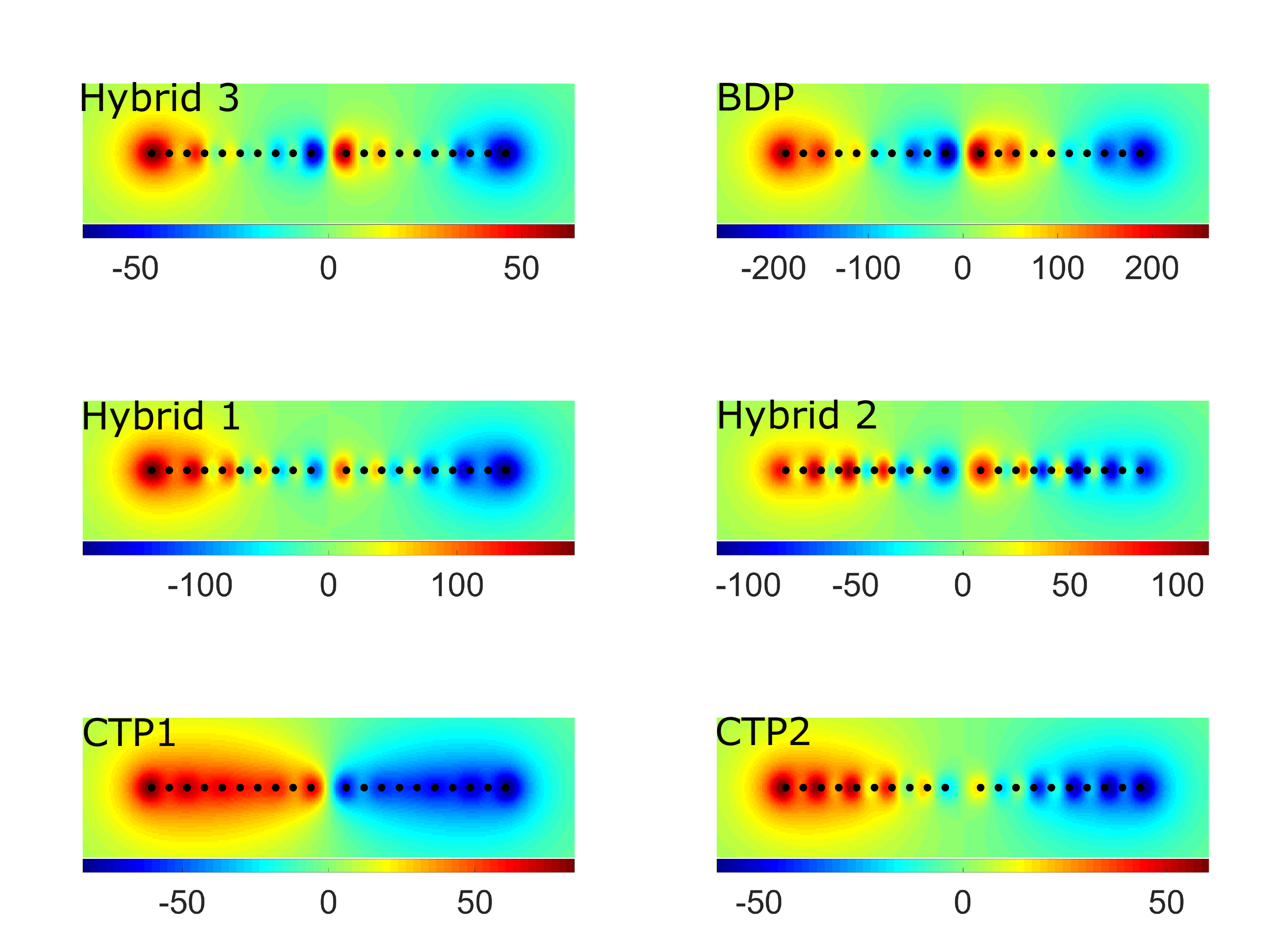}
\caption{The induced potential divided by the perturbing electric field strength for the 6 main tunnelling modes. `Hybrid 1' corresponds to the lowest energy and `Hybrid 3' to the largest energy hybrid modes shown in figure 5 in the main text. } \label{fig:Gap_Plasmons_Induced_Potential}
\end{figure}

\section{Transverse gap plasmons}\label{sec:Transverse_gap_plasmons}
In figure \ref{Sup_figure_4} we show the electronic spectra for a \ce{Na20} chain with a gap opened up in the center. We observe some interference effects between the TC and TE modes at the gap sizes considered and consequently some new modes are setup in the energy gap between the TC and TE modes (which are approximately at the same energy for the \ce{Na10} and \ce{Na20} chain).

\section{Induced potential of gap plasmons}\label{sec:ind_pot_gap_plasmons}
The induced potential can be calculated from the induced density by solving Poisson's equation $\nabla^2 (\delta \phi(\bm{r},\omega)) = e \frac{\delta n(\bm{r},\omega)}{\epsilon_0}$, the results of this calculation for the 6 major low-energy tunnelling modes is shown in figure \ref{fig:Gap_Plasmons_Induced_Potential}. As well as allowing the induced electric field to be calculated, it is also useful for identification of excitations. We find that the CTP1 mode is most CTP-like and the modes become increasingly BDP-like with increasing energy.

\bibliography{Bib}

\end{document}